\documentclass[fleqn,usenatbib]{mnras}

\usepackage{newtxtext,newtxmath}

\usepackage[T1]{fontenc}

\DeclareRobustCommand{\VAN}[3]{#2}
\let\VANthebibliography\thebibliography
\def\thebibliography{\DeclareRobustCommand{\VAN}[3]{##3}\VANthebibliography}

\usepackage{graphicx}	
\usepackage{amsmath}	
\usepackage{newtxtext,newtxmath}

\title[Exoplanet Detection using Machine Learning]{Exoplanet Detection using Machine Learning}

\author[A. Malik et al.]{
Abhishek Malik$^{1}$\thanks{E-mail: a.malik@usm.lmu.de},
Benjamin P. Moster$^{1}$, Christian Obermeier$^{1}$
\\
$^{1}$Universit\"ats-Sternwarte, Ludwig-Maximilians-Universit\"at M\"unchen, Scheinerstr. 1, 81679 M\"unchen, Germany\\
}

\pubyear{2021}

\begin{document}
\label{firstpage}
\pagerange{\pageref{firstpage}--\pageref{lastpage}}
\maketitle

\begin{abstract}
We introduce a new machine learning based technique to detect exoplanets using the transit method. Machine learning and deep learning techniques have proven to be broadly applicable in various scientific research areas. We aim to exploit some of these methods to improve the conventional algorithm based approaches presently used in astrophysics to detect exoplanets. Using the time-series analysis library \emph{TSFresh} to analyse light curves, we extracted 789 features from each curve, which capture the information about the characteristics of a light curve. We then used these features to train a gradient boosting classifier using the machine learning tool \emph{lightgbm}. This approach was tested on simulated data, which showed that is more effective than the conventional box least squares fitting (BLS) method. We further found that our method produced comparable results to existing state-of-the-art deep learning models, while being much more computationally efficient and without needing folded and secondary views of the light curves. For Kepler data, the method is able to predict a planet with an AUC of 0.948, so that 94.8 per cent of the true planet signals are ranked higher than non-planet signals. The resulting recall is 0.96, so that 96 per cent of real planets are classified as planets. For the Transiting Exoplanet Survey Satellite (TESS) data, we found our method can classify light curves with an accuracy of 0.98, and is able to identify planets with a recall of 0.82 at a precision of 0.63.
\end{abstract}



\section{Introduction}

Since the discovery of the first exoplanets \citep{1992Natur.355..145W, 1995Natur.378..355M}, the field of planet detection has become a major research area in astrophysics. To date, over 4,000 exoplanets have been discovered with various techniques, such as the radial velocity method \citep{1988ApJ...331..902C, 1989Natur.339...38L}, astrometry \citep{2002ApJ...581L.115B, 2010AJ....140.1657M}, direct imaging \citep{2004A&A...425L..29C, 2009ApJ...707L.123T}, and gravirational microlensing \citep{1991ApJ...374L..37M, 2006Natur.439..437B}. Yet, the majority of confirmed exoplanets have been discovered with the transit method \citep{2000ApJ...529L..45C, 2001A&A...375L..27N}.

Depending on the position of the observer, a planet may move in front of its host star blocking a part of the star's light, which causes the observed visual brightness of the star to drop by a small amount, depending on the relative sizes of the star and the planet. For the transit method, the brightness of a star is thus continually observed, whereupon `dips' in the obtained light curve are sought. One drawback of this method is that transits can only be observed when the planet's orbit is perfectly aligned from the location of the observer. Another issue is the high rate of false detections, e.g. through eclipsing binary systems or transits by planet sized stars, so that discoveries need to be confirmed with alternative methods. 


NASA's Kepler mission \citep{2010Sci...327..977B}, and its second survey programme K2 \citep{2014PASP..126..398H}, were a major step forward in the creation of a vast catalogue of exoplanet systems, which may give insights into the formation process of planets and the abundance of potentially habitable Earth-like analogues. Although Kepler was designed as a statistical mission to investigate the frequency of Earth-size exoplanets in or near habitable zones, most early results focused on individual systems \citep[e.g.][]{2012ApJ...747..144M, 2011ApJ...729...27B}, resulting in rather homogeneous catalogues. Later studies shifted their focus towards the statistics of the exoplanet population \citep[e.g.][]{2013ApJ...766...81F,2015ApJ...809....8B}. However, the validation process of the candidates discovered in the Kepler mission is still an ongoing process, as ruling out false-positive detections is time-consuming and new candidates are still being discovered.

In April 2018, NASA launched the Transiting Exoplanet Survey Satellite \citep[TESS;][]{2015JATIS...1a4003R} as a successor of Kepler with the primary objective to survey some of the brightest stars. It covers 85 per cent of the sky -- an area that is 400 times larger than that covered by Kepler -- and produces around one million light curves per month. TESS observes stars that are $30-100$ times brighter than those selected by the Kepler mission. Consequently, it is possible to identify targets that are far easier to follow up for detailed observation with other space-based and ground-based telescopes. Some of the early results of TESS include the discovery of a Super-Earth in the Pi Mensae system \citep{2018ApJ...868L..39H}, an ultra-short period planet orbiting a red dwarf  \citep{2019ApJ...871L..24V}, and an Earth-sized exoplanet in the habitable-zone \citep{2020AJ....160..116G}.

\begin{figure}
	\includegraphics[width=\columnwidth]{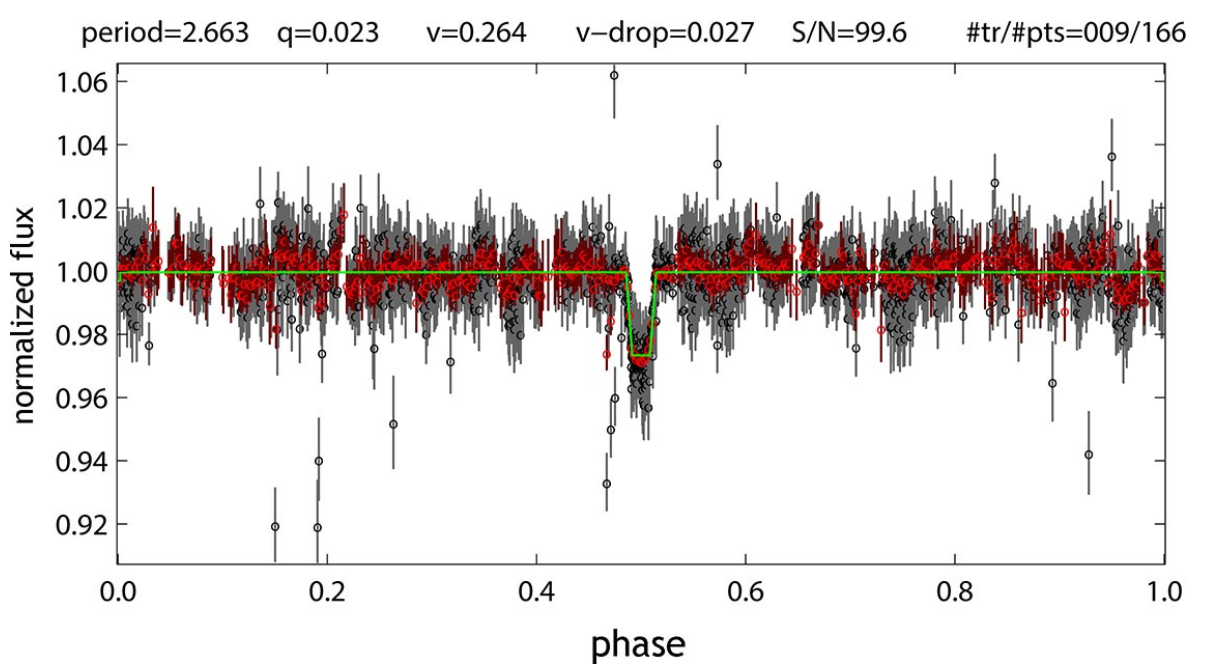}
    \caption{Phase-folded light curve of a K dwarf being orbited by a hot Jupiter candidate, observed by the Pan-Planets survey. The Box-fitting Least Squares (BLS) method has been used to fit the light curve. Shown at the top are the resulting parameters: period (days), transit duration q (in units of phase), transit v shape (0 corresponds to a box, 1 to a V), transit light drop, signal-to-noise (S/N) and number of transits/number of points in the transits. Binned data points are shown in red, while the green line shows the best-fit model. See \citet{2016A&A...587A..49O} for more details.}
    \label{fig:BLS}
\end{figure}

The first step in analysing observed light curves is to search for periodic signals that may be consistent with transiting planets, so-called ``threshold crossing events'' (TCEs). These TCEs are then inspected to eliminate erroneously selected non-signals (e.g. instrumental noise or astrophysical variability), either manually by humans, or in an automated way \citep[e.g.][]{coughlin2016planetary}. A typical phase-folded light curve from the Pan-Planets survey \citep{2016A&A...587A..49O} is shown in Figure \ref{fig:BLS}.

One of the most commonly used methods to detect exoplanet in light curves is Box-fitting Least Squares \citep[BLS;][]{2002A&A...391..369K}, as it is very effective in detecting periodical signals that can be approximated by a two level system such as transits. In practice, a box model is fit to the light curve, which yields parameters such as the period, the transit duration, and the transit light drop. The green line in Figure \ref{fig:BLS} shows the BLS fit to the light curve \citep{thesis_christian}. Cases with a seemingly good fit can then be manually reviewed. However, BLS is limited in terms of signal-to-noise and data cadence, and is vulnerable to false-positive detections created by cosmic rays, random noise patterns, or stellar variability.

Kepler, TESS, and other similar surveys still rely on manual analysis \citep[see][]{yu2019} for a good overview of this process). For a typical TESS sector, usually a group of experts manually eliminate obvious false positive cases, a process which alone can take a few days. From the remaining cases, each one has to be viewed by at least 3 experts. This kind of procedure can lead to disagreements on a particular case, as even experts might not maintain the same definition for the classification. A difference of opinion may arise in a team of experts on a particular case due to external factors like the way a case is presented, other TCEs viewed recently, or something as little as the time of day or their mood \citep[c.f.][]{coughlin2016planetary}.

For these reasons, systems are needed that can reliably and repeatably select the most important planet candidates for us, which can then be manually reviewed for confirmation at a later stage. Over the past years, there was a growing interest in building automatic vetting systems with machine learning methods, as they provide computers the ability to learn from data without being explicitly programmed, and are instead able to detect patterns automatically. Some of the notable earliest attempts include the Robovetter \citep{robovetter}, the Autovetter project \citep{mccauliff2015automatic} and SIDRA \citep{mislis2016sidra}, where tree-based models trained for vetting. However, the research moved away from classical machine learning methods recently, and deep learning methods became the new focus. Among the most notable work in this area was done by \citet{shallue2018}. They introduced a novel deep learning architecture, \emph{Astronet}, which produced very accurate results for the Kepler data. Their approach and their model architecture was adapted and applied to data from several different surveys such as Kepler's K2 mission \citep[]{shallue_2nd}, next generation transit survey \citep[NGTS;][]{ngts} and TESS data \citep{yu2019}.

Since the induction by \citet{shallue2018}, the community has moved on to deep learning, as deep learning methods tend to produce better results than classical machine learning approaches, especially for more complex problems or data types. However, deep learning models are usually computationally expensive and require large amounts of data to train properly. Often they are prone to overfitting the data if not regularised sufficiently. Moreover, it is usually difficult to understand, which input data (features) have been important to derive the results, which limits the physical interpretability. Therefore, in some cases, these models can be superfluous for the given problem and simpler or less computationally expensive approaches can perform equally well or even better, and can allow for a better understanding of the underlying problem.
On the other hand, it is also important to sometimes move away from the currently best performing methods and explore new alternatives. With this motivation, in this paper we propose a new direction to approach the problem of finding exoplanets in light curve data using classical machine learning.

This paper is organised in five sections. Section 2 contains details about our methodology, specifically data preparation, feature extraction and model training. Section 3 explains the results achieved on simulated, Kepler and TESS data. We also compare our results with some of the currently best performing models. In section 4, we compare the main differences how our model differs from the other deep learning-based methods being used in the area. We also provide details of our vetting tool which can be used to make real time inferences. Finally, we conclude and discuss future steps in section 5.

\begin{figure*}
	\includegraphics[width=0.95\textwidth]{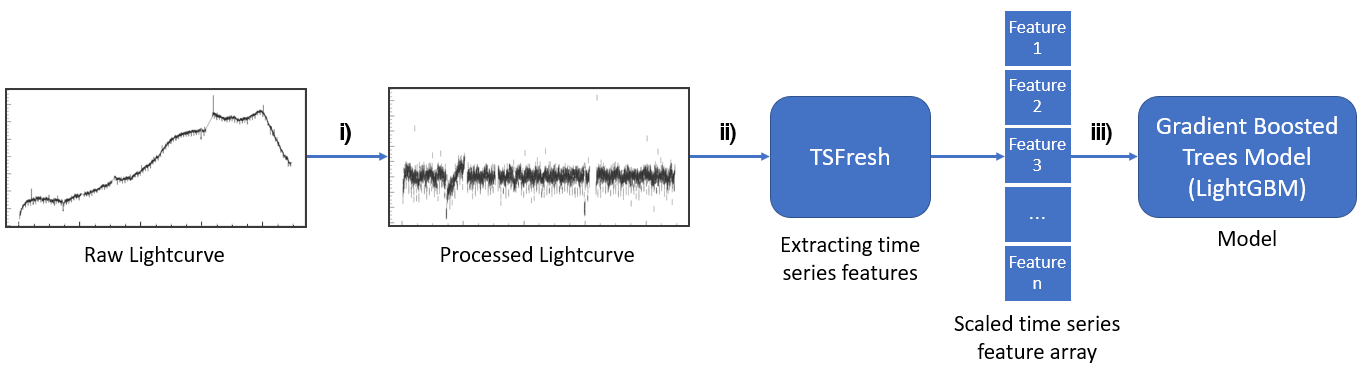}
    \caption{Workflow of our method starting from raw light curves to performing inference on the sample. I the first step (i) the raw light curves are processed to remove low frequency variability and noise, and to sample the curve uniformly in time. In the second step (ii) features are extracted from the light curves and organised in a feature array. In the final third step (iii) the feature arrays are used as the input for a classification algorithm (gradient boosted trees) and the model is trained.}
    \label{fig:workflow}
\end{figure*}

\section{Methods \label{method}}
Machine Learning methods are widely used in scientific research areas to build classifiers i.e. an algorithm which separates data into two or multiple classes. In our case we are building a binary classifier which will separate each time-series photometry, a so-called light curve, into the classes `planet candidate' and `non-candidate'. The state-of-the-art machine learning method for planet detection presented by \citet{shallue2018} utilises deep learning, a class of machine learning methods based on artificial neural networks. Our method on the other hand is based on classical machine learning. One essential difference between our approach and deep learning is that the latter is able to extract features automatically, while for classical machine learning the features have to be calculated beforehand and provided as input to the model. To this end, we use Time Series Feature extraction based on scalable hypothesis tests \citep[\emph{TSFresh};][]{tsfresh}, a python-based library for feature extraction.
The idea behind the method was inspired by methodology used in time-series prediction projects employing feature engineering libraries. \emph{TSFresh} is also used in projects like machine fault prediction \citep{faultprediction}, diagnosing Alzheimer’s disease \citep{alzheimer}, identifying epileptic seizures in electroencephalography signals \citep{seizures}, earthquake prediction \citep{earthquakes}, and time series forecasting for business applications \citep{businessforecasting}. Light curves are essentially a time series and so these tools can directly be used for our case.

We trained and tested our model on three kinds of data sets. The first stage used simulated data with K2 photometry as a baseline with additional injected transits. We then trained it on Kepler and finally TESS photometry. Each stage is split into three parts:
\begin{enumerate}
    \item processing and labelling the training data;
    \item extracting features from each light curve using \emph{TSFresh};
    \item model training.
\end{enumerate}
This workflow is shown in Figure \ref{fig:workflow}. More details on the individual steps are given in the following sections.

\subsection{Preparing and Labelling Training Data}

\subsubsection{Simulated Data}
\label{sec:simulated_data}
We obtained the K2 photometry from the Mikulski Archive for Space Telescopes (MAST) and used the calibration by \citet{2014PASP..126..948V}. While the processing of \citet{2014PASP..126..948V} already removed the vast majority of systematic effects, we further cleaned up the data by identifying and removing remaining cosmic ray artefacts and creating a new noise model. If a point is an outlier with respect to its neighbours, i.e. if it is more than $5\sigma$ above its previous and following point, it was assumed to be a cosmic ray artefact. Then we removed the stellar variability that is common, depending on stellar type, with an iterative process. We smoothed the data, binned it, fit cubic splines to the binned data, clipped negative $3\sigma$ outliers and iterated this process until it converged. 

\begin{figure*}
	\includegraphics[width=0.95\textwidth]{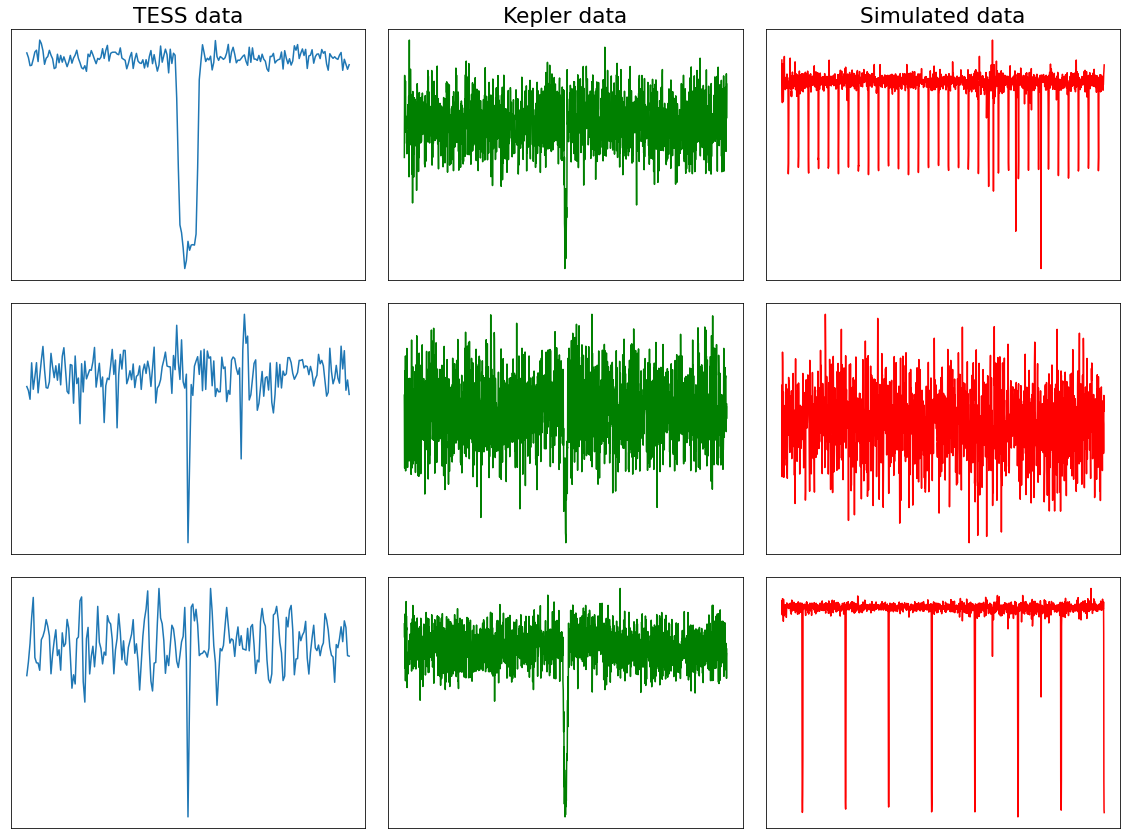}
    \caption{Sample light curves containing transit signals scaled to arbitrary units. All of these curves are taken from our test set of the respective data set. It can be observed that the TESS (blue) and Kepler (green) data are a lot more noisy and likely to contain a lower number of transits. Simulated data (red) on the other hand is cleaner, as all the curves with S/N over 12 were removed from the data set before the planet signals were injected. However, some of the injected transits were generated with low S/N values to imitate the realistic scenarios. The second light curve of simulated data is one such case.}
    \label{fig:sample_lc}
\end{figure*}

Transit signals were randomly injected in half of the processed light curves in an approach similar to that introduced by \citet{2016A&A...587A..49O}. We removed known planet systems from the data sample beforehand and then performed a blind search for eclipsing binary systems based on our implementation of the BLS method. As the K2 light curves sometimes had gaps, we interpolated between those gaps to sample the time series uniformly. Any light curve with a detected signal-to-noise ratio of more than 12 was removed from the sample. We then created a randomly filled space of orbital periods, stellar limb darkening and radii, planet radii and orbit inclinations. An example of the resulting simulated planet systems can be seen in Figure \ref{fig:sample_lc}.

Each light curve with injected planets was labelled as class `1' and remaining curves were categorised as class `0'. These labels were then used to train our classifier with the objective of identifying the injected transit signals. In total, our training set consisted of 7,873 light curves out of which 3,937 belong to class `1' (planet candidate). This data set was then divided into training-validation set for 10-fold cross validation (90 per cent) and test set (10 per cent).

\subsubsection{Kepler Data}
\label{sec:keplerdata}
We used the publicly available data set that was employed in the work of \citet{shallue2018}; the details about data processing can be found in section 3.2 of their paper. These light curves were produced by the Kepler pipeline \citep{jenkins2010b}, where each light curve contains around 70,000 data points equally spaced out at the interval of 29.4 minutes. The light curves were flattened and outliers were removed as described in the previous section. The labels for the curves were taken from the Autovetter Planet Candidate Catalog \citep{catanzarite2015autovetter}. The catalog is divided into four light curve categories: planet candidate (PC), astrophysical false positive (AFP),  non-transiting phenomenon (NTP) and unknown (UNK). All UNK light curves were removed from the data set and all PC light curves were given class label 1 (`planet candidate'). Both the remaining cases were assigned to class 0 (`non-candidate'). In total, there are 3,600 PC light curves and 12,137 non-PC light curves. The data is already divided into training set (80 per cent), validation set (10 per cent), and test set (10 per cent). We used the same test set for our model performance as \citet{shallue2018} in order to compare our model performance. Apart from this, we combined their training and validation set into a training-validation set for 10-fold cross validation.

\subsubsection{TESS Data}
\label{sec:TESSdata}
For the TESS data, we again used publicly available data provided by \citet{yu2019} for their model \emph{AstroNet-Vetting}. Details about their light curve processing can be found in the section 2.1, where they used the MIT Quick Look Pipeline \citep[QLP;][]{2020RNAAS...4..204H} for processing the light curves. The QLP is designed to process the full-frame images (FFI) and produces light curves using its internal calibrated images. The labelling was done by visually inspecting each light curve, categorising them into 3 classes: Planet Candidates (PC), Eclipsing Binary (EB), and Junk (J), where junk signals were the cases with stellar variability and instrumental noise. The data set consists of 2,154 EB, 13,805 J and 492 PC signals, where PCs were labelled as class 1 (`planet candidate') and everything else was labelled as class 0 (`non-candidate'). These data are also already divided into training set (80 per cent), validation set (10 per cent) and test set (10 per cent). Like the previous case, we used the same test set and combined the training and validation set into a training-validation set for 10-fold cross validation.

Some of the sample light curves containing transit signals are shown in figure \ref{fig:sample_lc}. All of these curves are taken from our test set of the respective data set. It can be seen that the TESS (shown in blue) and Kepler (shown in green) data are a lot more noisy and likely to contain a lower number of transits. On the other hand, the simulated curves are cleaner as all the curves with S/N over 12 were removed from the data set before the planet signals were injected. However, some of the injected transits such as the second light curve of the simulated data were generated with low S/N values to imitate realistic scenarios.

\subsection{Feature Extraction}
\label{sec:featureextraction}
Having processed the light curves, the next step is to extract features, which capture information about the characteristics of each light curve and are used as the input for our classification model both for training and inference. We used the python framework \emph{TSFresh} \citep{tsfresh} to extract features that are typical for time series such as light curves, e.g. the absolute energy of the time series (the sum over the squared values), the number of values that are higher than the mean, and the coefficients of the one-dimensional discrete Fourier transform. For all used data sets, the light curves are resampled to the frequency of 1 hour,  i.e. all light curves are reformatted into windows of 1 hour to ensure that data points are uniformly distributed in time, a process commonly referred to as `resampling'. Although this resampling process is not mandatory, it is a standard practice and proven to produce a better representation of the data, leading to faster convergence. In the resampling process, if multiple data points are inside one window, they are summed up and missing data points on the curves are interpolated. The frequency of one hour is chosen to ensure minimum data loss while not increasing the data size significantly. 

These resampled time series can now be directly used with multiple time-series analysis tools. We used \emph{TSFresh}'s efficient feature extraction setting which extracted around 790 generalised time series features\footnote{Details of these features can be found at: \url{https://tsfresh.readthedocs.io/en/latest/text/list_of_features.html}}. After the feature extraction process, we implemented some standard data pre-processing steps for machine learning. We removed all irrelevant features, i.e. those with values that were constant throughout the data set, and imputed missing values by interpolating. Lastly, we scaled the whole data set using a robust scaler. This data set was then used for training a tree-based classifier.

\subsection{Model Training}
\label{sec:training}

Our model is a binary classifier, i.e. it classifies every light curve into two classes, either `planet candidate' or `non-candidate'. We employed a gradient boosted tree (GBT) model using the popular machine learning framework \emph{lightGBM} \citep{lightgbm} for our classifier. A GBT model is an ensemble of decision trees which are trained in sequence. In each iteration, the GBT model is trained to reproduce the results of the decision trees by fitting the negative gradients \citep[also known as residual errors; see][]{friedman2001greedy}.

Moreover, we used a 10-fold cross validation (CV) during training. This means the training set is split into 10 smaller sets and the model is trained using 9 sets at a time. The remaining last set is used to evaluate the model performance. This is done iteratively until the model has been evaluated on all 10 sets. The final performance is then the average of all the values computed in the iteration. 

We use the following four metrics to evaluate our results:
\begin{itemize}
\item Accuracy: the fraction of samples correctly classified, including both `planet candidates' and `non-candidates'.
\item Precision: the fraction of predicted `planet candidates' that are true planets, i.e. the accuracy of the `planet candidate' predictions. 
\item Recall: ratio of the true planets correctly detected by the classifier as `planet candidate', i.e. what fraction of true planets is recovered (also called ``true positive rate'').
\item AUC: the Area Under the receiver operating characteristic (ROC) Curve, called the AUC, gives the probability that a randomly chosen light curve with a true planet is ranked higher than a randomly chosen light curve without a planet.
\end{itemize}

The accuracy is generally not a proper metric to judge the performance of the planet detection algorithm, since most of the exoplanet detection data sets are unbalanced, i.e. usually there are significantly more light curves without any planet signal than cases with a planet. For instance, in the TESS data set only 3 per cent of the light curves are true planet candidates. If a classifier simply predicts `non-candidate' for all light curves it will still lead to a `high' accuracy of 97 per cent, even though it has not learned from the data and cannot identify any planet candidates.

While a high precision ensures that most of the predicted `planet candidates' are true planet candidates, it is also not a very useful metric, as a trivial way to achieve a high precision is to make only few `planet candidate' predictions and ensure they are correct. Consequently, many possible planet candidates would be missed. As we would rather allow for a higher number of false positives than losing possible planet signals, the recall is a much more important metric to assess the performance of our algorithm.

A model with a high recall may lead to a lower precision and vice versa, which is commonly known as the precision-recall trade-off. Classification decisions are based on a decision threshold, which is another hyperparameter of the model. If the predicted model probability surpasses this threshold, the light curve is classified as `planet-candidate'. For a typical classification problem a threshold of $p_\mathrm{threshold}=0.5$ is chosen, so that light curves with a predicted probability of more than 0.5 are considered `planet candidates'. If a higher decision threshold is chosen, the model will have a higher precision but a lower recall, and vice-versa, so it can be optimized to yield a high recall.

Finally, the AUC is independent of the decision threshold, as it gives the integral over all possible decision thresholds. Therefore, it can be used to optimise the remaining hyperparameters first to get a model that performs well for any threshold. In the second step, the threshold can then be chosen to achieve a high recall. We argue that out of the above four metrics AUC and recall are the most important as the main imperative is to design a method that is able to recover a high percentage of light curves containing planet signals. In other words, we aim to optimise our model for not missing light curves that contain a true planet signal.

Due to these reasons, we first optimised all hyperparameters of our model to maximize the AUC, and then optimised the decision threshold to obtain a high recall while maintaining precision as high as possible. In practice, once the all hyperparameters except for the threshold are fixed, we check whether it is possible to get both precision and recall in the upper percentile ($> 90$ per cent), and if so select the highest possible recall while maintaining a precision in that percentile. If this is not possible, we allow the precision to drop to the next percentile and select the highest possible recall.

\section{Results}

\subsection{Simulated data}

\begin{figure}
	\includegraphics[width=\columnwidth]{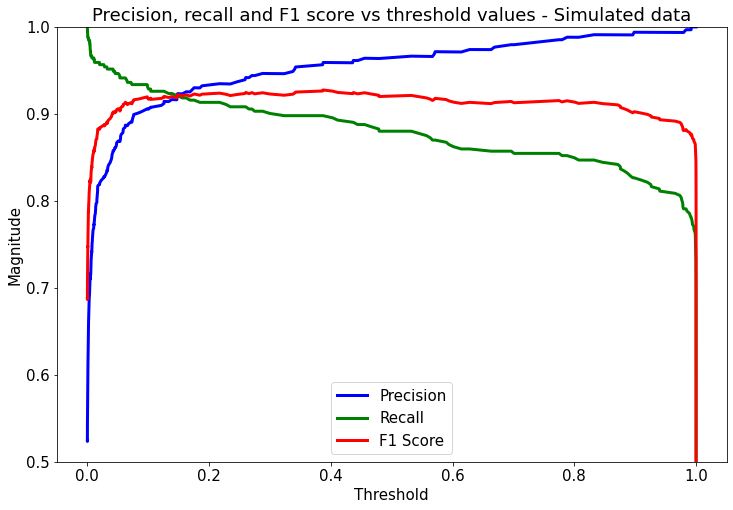}
    \caption{Precision, recall and F1 score (harmonic mean of precision and recall) as function of the decision threshold for the test set of our simulated data. The typical threshold of 0.5 for classification problems can be adapted to increase or decrease the model sensitivity. For the simulated planet data set, choosing a threshold of 0.13 results in a recall of 0.94 and a precision of 0.92. As the objective is to retrieve the maximum number of possible light curves containing transit signals, this result is preferred over the ones produced by a default threshold of 0.5. The AUC of the model is 0.92.}
    \label{fig:sim_threshold}
\end{figure}

In order to attain a proof of concept the method was first applied to our simulated data that was obtained as described in section \ref{sec:simulated_data} and consists of 7,873 light curves. Extracting features for each of the light curves and using them as inputs for the GBTs, the model outputs the probability for a light curve to contain a transit. Given a decision threshold, the light curve is then either classified as `planet-candidate', or `non-candidate'. Figure \ref{fig:sim_threshold} shows the precision, recall, and the F1 score (the harmonic mean between precision and recall), as a function of the decision threshold. All curves have been computed with the model that was optimized for a high AUC as described above. Choosing a decision threshold of $p_\mathrm{threshold}=0.13$, we get a recall of 0.94 and precision of 0.92. This result is preferred over the one with the standard threshold of 0.5 as for planet detection, it is preferred to extract the maximum possible light curves at the cost of a minor increase in false positives. The results on our validation set are shown in the Table \ref{tab:simulated}.

\begin{table}
	\centering
	\caption{Results on simulated data}
	\label{tab:simulated}
	\begin{tabular}{lccr} 
		\hline
		AUC & Recall & Precision & Accuracy\\
		\hline
		0.92 & 0.94 & 0.92 & 0.91\\
		\hline
	\end{tabular}
\end{table}

We then used the BLS algorithm on our simulated data set and were able to detect only around 84 per cent of the true planets as opposed to 94 per cent with our machine learning based method. Overall our machine learning model failed to identify around 63 (out of 970) cases, while BLS failed to detect 155 cases. We also found an overlap in the cases that were not detected by our method and BLS. We investigated these cases manually and found that in this sample the transits were injected randomly with random parameters which resulted in many non-detectable transits such as:
\begin{itemize}
    \item Cases with a relatively low inclination angle for the given star, which resulted in very small planet signals.
    \item Cases with low S/N ratio where the injected transit signal was weaker than the noise, and hence it was not detected either by our method or by BLS.
\end{itemize}

The majority of these cases had a signal-to-noise ratio of $\mathrm{S/N} < 12$, so that the signal was simply too weak.
These results demonstrate that our machine learning based method can detect planets more accurately and efficiently than BLS, especially in cases with low signal-to-noise ratio.

Similar results were also shown by \citet{AIoverbls}, who compared various machine learning methods with BLS. Their machine learning models were able to detect planet signals with higher signals and much lower false positive rate compared to BLS.
Therefore, our next step was to apply our methodology on realistic data sets, i.e. the Kepler and TESS light curves.

\subsection{Kepler data}

\begin{figure}
	\includegraphics[width=\columnwidth]{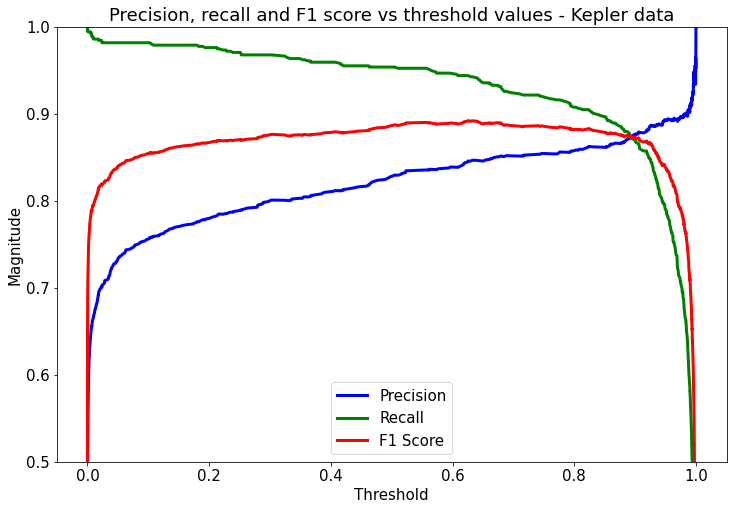}
    \caption{As Figure \ref{fig:sim_threshold}, but for the Kepler data set. Choosing a threshold of 0.46 results in a recall of 0.96 and a precision of 0.82. The AUC of the model is 0.948.}
    \label{fig:kepler_Threshold}
\end{figure}

Using the Kepler data set, \citet{shallue2018} produced very promising results with a deep learning model. To date, these are the best current results on this data set. They shared their training dataset publicly, which we used to evaluate our method, and we used their results as a benchmark for our results. The data set consists of 15,737 light curves that were labelled into 3 classes by the Autovetter and combined into a `planet candidate' class and a `non-candidate' class, as described in section \ref{sec:keplerdata}. We again extracted features from the light curves with \emph{TSFresh} and used them along with the class labels to train a GBT model. Optimizing the hyperparameters, we found the best model to yield an AUC of 0.948. The resulting precision, recall, and F1 score as a function of the decision threshold are shown in Figure \ref{fig:kepler_Threshold}.

As discussed in section \ref{sec:training}, we decided to optimize the threshold to obtain the highest possible recall, while maintaining a good precision. Choosing a threshold of $p_\mathrm{threshold}=0.46$ results in a recall of 0.96 and a precision of 0.82. A comparison of the results of our method with the results by \citet{shallue2018} is summarised in Table \ref{tab:kelper}. It can be seen that the results of both the methods are comparable. While our model has a somewhat lower AUC, it has a slightly higher recall, albeit at a considerably lower precision. However, as discussed before, we specifically aimed at getting a high recall instead of precision, as on major criterion is to identify as many planet candidates as possible.

\begin{table}
	\centering
	\caption{Results on Kepler data}
	\label{tab:kelper}
	\begin{tabular}{lccr} 
		\hline
		Type & AUC & Recall & Precision\\
		\hline
		\citet{shallue2018} & 0.988 & 0.95 & 0.93\\
		Our method & 0.948 & 0.96 & 0.82\\
		\hline
	\end{tabular}
\end{table}

\begin{figure}
	\includegraphics[width=\columnwidth]{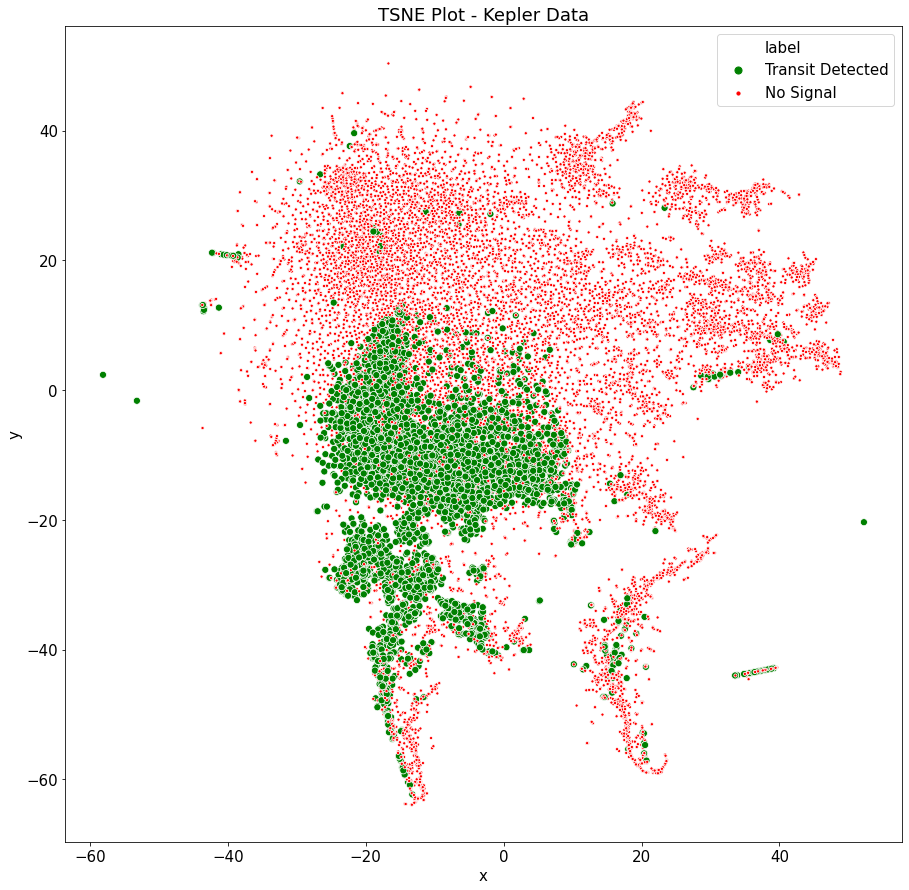}
    \caption{t-SNE visualisation of the Kepler data. Light curves with planet transit signal(s) are shown in green. It can be observed that these curves are clustered in one region of the space.}
    \label{fig:tsne_kepler}
\end{figure}

To further analyse the Kepler data set and to visualize how different light curves cluster, we embedded the data in a low-dimensional space of two dimensions with the t-distributed stochastic neighbor embedding \citep[t-SNE;][]{tsne} algorithm. To this end, we extracted time series features from each light curves as described in section \ref{sec:featureextraction} . These features were then scaled using a robust scaler, and any missing values were filled using the mean of that column, which resulted in around 700 processed features. The top 70 per cent of those features were selected, based on the feature importance found by the GBT model (as using only the top 70 per cent of the features resulted in an optimized AUC), and used as the input for the t-SNE algorithm for dimensionality reduction. The results are presented in the Figure \ref{fig:tsne_kepler}.

As the t-SNE algorithm retains the local structure of the data, similar light curves are modelled by nearby points with high probability, i.e. the points closer in the 2D representation are also closer in the high dimensional space. However, a point far away in the 2D space might not be far away in high dimensional space. As Figure \ref{fig:tsne_kepler} shows, most of the light curves containing a transit are clustered in one region of the plot, which further validates our pre-processing and feature extraction process. Broadly there are 3 main clusters in the plot, and samples from these clusters can be analysed to understand how these clusters vary from each other. However, we left this analysis for future investigations.

\subsection{TESS data}

TESS light curves are in many way different from the Kepler data set. Kepler observed a fixed field of view in its 4 year timeline and K2 observed each of its 19 campaigns for 80 days. On the other hand, TESS observed each sector for 27 days. Longer baselines contain more details and lead to a much higher signal to noise in terms of detections, hence the extracted features are also more rich in information. 
The short time span of TESS implies that each light curve (and features extracted from it) tends to contain less data points with transits. Consequently it becomes more difficult for a classifier to differentiate a case with a planet signal from a case without it. Moreover, the short length of the light curves makes the presence of multiple-periodic transit signals less likely to be observed. 
For longer exoplanet transits, this problem is further compounded if only a single transit is recorded. In contrast, multiple transiting planets in the Kepler data may lead to an automatic confirmation due to the low probability of any fitting false-positive scenario.

\begin{figure}
	\includegraphics[width=\columnwidth]{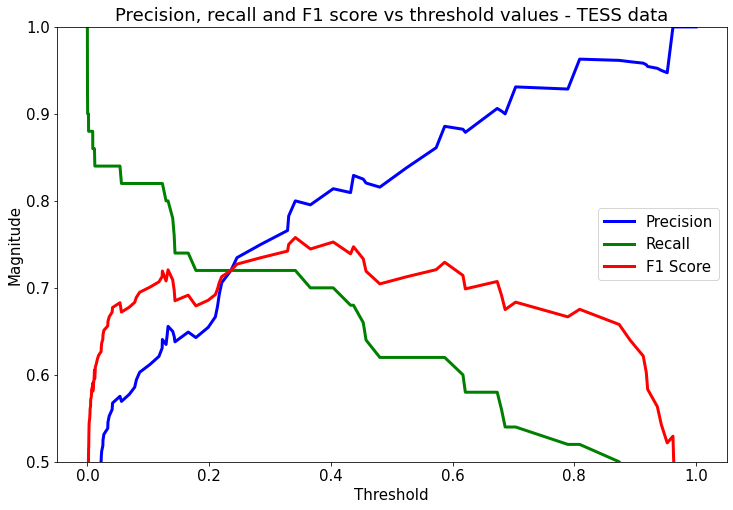}
    \caption{As Figure \ref{fig:sim_threshold}, but for the TESS data set. With a threshold of 0.12 the recall is 0.82 and the precision is 0.63. The AUC of the model is 0.80.}
    \label{fig:TESS_Threshold}
\end{figure}

The first machine leaning classifier to be trained and tested on real TESS data was introduced by \citet{yu2019}. Their model is an adapted version of \citet{shallue2018} as shown in Figure 2 of their paper. We again trained our model on the data set publicly shared by \citet{yu2019} and used their results as benchmark. The data consists of 16,500 light curves, with only 490 planet candidates. Applying \emph{TSFresh} and a GBT model, we optimized the hyperparameters to get an AUC of 0.81. The resulting precision, recall, and F1 score as a function of the decision threshold are shown in Figure \ref{fig:TESS_Threshold}. We again optimized the decision threshold to get a high recall, which resulted in a threshold of $p_\mathrm{threshold}=0.12$, a recall of 0.82, and a precision of 0.63 for the `planet candidate' class. A comparison of the results of both the methods with default and optimal thresholds is shown in Tables \ref{tab:tess} and \ref{tab:tess_thresh}, respectively.

\begin{table}
	\centering
	\caption{Results on TESS data with default threshold}
	\label{tab:tess}
	\begin{tabular}{lccr} 
		\hline
		Type & AUC & Recall & Precision\\
		\hline
		\citet{yu2019} & 0.98 & 0.57 & 0.65\\
		Our method & 0.81 & 0.62 & 0.84\\
		\hline
	\end{tabular}
\end{table}

\begin{table}
	\centering
	\caption{Results on TESS data with optimized threshold}
	\label{tab:tess_thresh}
	\begin{tabular}{lccr} 
		\hline
		Type & AUC & Recall & Precision\\
		\hline
		\citet{yu2019} & 0.98 & 0.89 & 0.45\\
		Our method & 0.81 & 0.82 & 0.63\\
		\hline
	\end{tabular}
\end{table}

With the optimized threshold, our model was able to find 40 out of 49 curves with planet transit in our test set as opposed to 44 samples identified by \cite{yu2019} on this test set. On the other hand, the precision of our model is significantly higher, i.e. it will result in almost half of the false positives as encountered by \cite{yu2019}. With the default threshold, the deep learning method by \citet{yu2019} resulted in a much higher AUC than our model, but only had a recall of 0.57, so only about 57 per cent of all planets were detected by the model. On the other hand, our machine learning method is able to obtain a recall of 0.62.

Of course, a recall of 0.82 is still not ideal and there is a lot more work to be done before such system can be used in production, both in preparing the data set and tuning the hyperparameters. For planet detection with both the Kepler and the TESS data sets, class imbalance is one of the biggest problems the model is facing. Now that more and more planets are identified in the TESS data, more light curves with planets can be gathered and used to train machine learning models, which is expected to improve the results (see the discussion in section \ref{sec:discussion}). Even though our machine learning algorithm is not yet fit for production with the TESS data, it is certainly possible to use the model for reliably eliminating false positives, so that the amount of human vetting can be greatly reduced on cases that are a clear false positive.

\begin{figure}
	\includegraphics[width=\columnwidth]{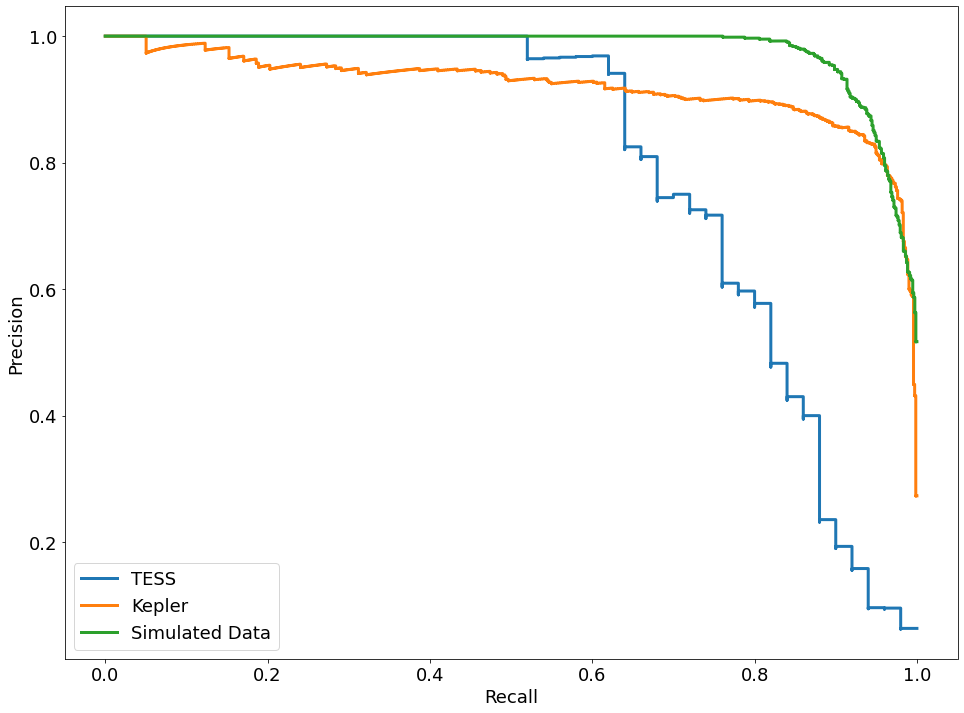}
    \caption{Precision as a function of recall for the test sets of TESS, Kepler and the simulated data, demonstrating the trade-off between precision and recall for different decision thresholds. The performance for the simulated and the Kepler data is similar, with a slight increase in the number of false positives for the Kepler data set, as it is more noisy. The performance for the TESS data set is worse as a result of a high class imbalance.}
    \label{fig:pr_curve}
\end{figure}

The performance for all three data sets is compared in Figure \ref{fig:pr_curve}, where we plot the precision as a function of recall for each data set. We can see that the performance for the simulated and the Kepler data is similar. There is a slight increase in the number of false positives in the Kepler data set, as it is more noisy than the simulated one. On the other hand, the performance for the TESS data set is significantly worse, with the high class imbalance being the primary cause of this.  With a more balanced data set, the performance for the TESS data is expected to approach the other two cases.

\section{Discussion}
\label{sec:discussion}

\subsection{Classical Machine Learning vs Deep Learning}

One of the primary differences between classical machine learning and deep learning methods is the fact that deep learning models are able to automatically extract useful features from raw data, while classical machine learning methods usually cannot deal with raw data and need extracted features. Consequently, the classical machine learning approach does not work well when problems become very complex, such as language translation or image classification, where instead deep learning is pushing new frontiers. However, classical machine learning methods should not be disregarded, as they are much more efficient and can still produce valuable results for several problems, such as classifying quasars \citep{2015A&A...584A..44C}, or identifying asteroids \citep{2017MNRAS.469.2024S}.

Our classical machine learning approach has the following advantages over state-of-the-art deep learning methods:
\begin{itemize}
    \item It can work with only a global view of the light curve, whereas deep learning additionally requires folded or secondary views.
    \item Training is less time consuming and takes less than 5 minutes to train on a 2 CPU system. This also indicates that it can be quickly adapted to a new data source, while deep learning can take more than 5 hours for training and much longer for tuning hyperparameters.
    \item The exact same model setup and code can be used for data from different sources such as Kepler, K2 and TESS. Hyperparameters need to be optimized only once, whereas deep learning almost always needs re-tuning when the data set changes.
    \item The most important features can be automatically identified, which allows for a better understanding of the data and the underlying physical processes. By contrast, deep learning models are usually black boxes, so that it is hard to interpret the results.
    \item No special hardware, such as a GPU, is required for training.
\end{itemize}
However, our approach has the following disadvantages compared to state-of-the-art deep learning methods:
\begin{itemize}
    \item Generally, the performance is lower compared to deep learning, which when properly trained can often achieve better results.
    \item Time series data such as the global view, and folded view can be directly used as an input for deep learning, while a classical machine learning model requires extracted features from the time series data.
\end{itemize}

\subsection{The Vetting Tool}

Machine learning methods are still relatively new in astrophysics and often met with scepticism. While there is a new paradigm of building classification techniques with machine learning methods, it still has a long way to go before being used in production pipelines. One major factor is the non-deterministic or black-box nature of such methods. Unlike conventional algorithms, the outcome of such methods cannot always be understood due to the non-linear and stochastic nature of such methods, so that it is harder for the user to explain the reason behind predictions made by the model. Even though these models can outperform conventional methods like BLS, they do not always make correct predictions which is another contributing factor for the scepticism. Hence, it is important to note that they cannot completely replace human vetting experts. However, if they are supported to be used alongside the domain knowledge, these techniques can help to automate processes like planet detection and enable us to deal with the growing data size in astronomy.

These tools are not widely used yet but for these systems to progress further, it is important to take feedback and expertise from the community on the methods. For this reason, we have developed a vetting tool where a few results from our model can be explored in an interactive way without any knowledge of the underlying machine learning method or the feature calculation technique. The tool is hosted on GitHub\footnote{\url{https://github.com/abhmalik/Exoplanet-Vetting-Tool}}.

The vetting tool allows the user to explore cases where the model assigns a high probability of planet candidacy. It allows the user to view the light curve in the global view and the folded view (also called local view). It further gives a list of features that were important in the model prediction for that case. It can be easily adapted to make model inference on any new light curve.

\subsection{Future steps}
With this paper, we tried to provide a new direction for a light and efficient automatic vetting system. However, there is a lot more that can be done in this direction.

Since our method is able to work with data from different sources such as Kepler, K2, and TESS, it is possible to construct a global classifier that can process light curves of any length from any data source directly. This would enable us to continue using the same model in case the data format changes, as happened for K2 when light curves had a time gap.

As we saw, the performance of our model was worse for the TESS data set compared to the other two data sets. The primary reason for this was the high class imbalance in TESS data, i.e. less than 3 per cent of the light curves contain transit signals, and most of them are just single transits. Now that more and more planets are being confirmed from TESS data, it will be possible to construct a data set with better class balance, which might help us to reach the performance levels we are reaching with the Kepler and the simulated data.

As shown in the t-SNE plot in Figure \ref{fig:tsne_kepler}, light curves are clustered in different regions. We can extract samples from these groups and analyse how the light curves in different group differ. This information can further be used to optimize the model performance for that particular data set.

Another possibility would be to forego the supervised learning approach and turn towards a semi-supervised approach. To this end the data would first have to be embedded in a lower-dimensional space (e.g. using an autoencoder, or principal component analysis). An unsupervised clustering algorithm could then find clusters of data in the second step. Finally, a limited amount of well-understood and labeled light curves could be used to label the individual clusters. If a new light curve would then be determined to be a member of a specific cluster, the cluster label can thus be used for the light curve. We leave such an approach for future work.

\section{Conclusions}

Machine learning methods have seen very active development in the last decade and now they are an essential part of our way of working. In fact, for everyone who interacts with a computer or smartphone, it is highly likely to interact with some sort of machine learning program.
They are also widely used in sciences for several use cases such as detecting diseases \citep{PMID:16834566} and finding string models in a string theory landscape \citep{2020arXiv200313339D}.
Within astronomy, there have been many applications such as classifying galaxy morphology \citep{2019MNRAS.482.1211W}, creating new large-scale structure samples without long simulations \citep{2019ComAC...6....1M}, and identifying gravitational waves \citep{2018PhLB..778...64G} and gravitational lenses \citep{2018MNRAS.473.3895L}.

With new and advanced telescopes, data in astronomy are growing at a fast pace. Conventional methods that involve human judgements are not efficient and prone to variability depending on the investigating expert. For example, the commonly used method for exoplanet detection, BLS, produces large number of false positives in case of noisy data, which have to be reviewed manually.

In this paper, we proposed a novel planet detection method based on classical machine learning. Our method consists of automatically extracting time series features from light curves which are then used as input to a gradient boosted trees model. We were able to demonstrate that with our machine learning method, we could identify light curves with planet signals more accurately as compared to BLS while significantly reducing the number of false positives. 

Pioneering work in the area has been done by \citet{shallue2018} using deep learning. Although deep learning methods often outperform classical machine learning approaches, specifically in complex problems such as machine translation and object detection, they tend to be harder to optimize and usually require a long training time on special hardware (GPUs). Moreover, they typically rely on large sets of training data and are more difficult to hypertune as they have to navigate a much bigger parameter space as compared to classical machine learning models. Hence, in this work, we attempted to move away from the standard deep learning approach and introduced a new direction that provides a light-weight model.

Our approach consists of three main steps. First, the raw light curves were processed, so that systematic effects, cosmic rays, and stellar variability were removed to get flat curves with a uniform time sampling. In the second step, the processed light curves were then analysed by the \emph{TSFresh} library, which extracted features that are typical for time series (e.g. absolute energy and number of values above mean). Finally, the third step was to use these features to train a gradient boosted tree model to predict the class (`planet candidate' or `non-candidate') for each light curve. In this way, our model is able to distinguish important features from unimportant ones, and there is no need for manually selecting features. Moreover, our approach does not require special hardware like GPUs and it takes about 2 minutes to train our classical machine learning model on a dual core CPU system. This means it can be easily trained even on low end computers and can be revised quickly in case it is required (e.g. when new data is available).

We found that our machine learning method is able to outperform the classical method to identify planet signals, BLS. For our simulated data set, BLS was able to identify 84 per cent of true planets, while the machine learning model could identify 94 per cent. We also compared the performance of our results with other state-of-the-art models based on deep learning, and found that, for the Kepler data, our model is able to achieve comparable results (a recall of 0.96 at a precision of 0.82). For the TESS data our model is able to achieve a recall comparable to the deep learning models (82 vs 89 per cent), but at a precision that is significantly higher (81 per cent vs 44 per cent), so that fewer false positives have to be removed manually.

While these results are very encouraging, such systems are not yet robust enough to be broadly applicable. Even if a machine learning model works well in the development environment, it is prone to make mistakes on unseen data. Hence, such methods should be used alongside with human supervision. Nonetheless, at the current stage, these models can provide a very reliable system to rule out large number of false positives and can drastically reduce the number of cases requiring manual reviews.

\section*{Acknowledgements}

We are grateful to Arno Riffeser and Jana Steuer for useful discussions and comments. 
We would also like to thank Christopher Shallue and Liang Yu for publicly sharing their code and training data. This enabled us to quickly test our model on different data sets and provided a benchmark to compare our results.
We thank the developers of Lightkurve \citep{lightkurve}, NumPy \citep{walt2011numpy}, Matplotlib \citep{Hunter:2007}, Scikit-learn \citep{scikit-learn} and TensorFlow \citep{tensorflow2015-whitepaper} for their very useful free software. The Astrophysics Data Service (ADS) and arXiv preprint repository were used intensively in this work.
BPM acknowledges an Emmy Noether grant funded by the Deutsche Forschungsgemeinschaft (DFG, German Research Foundation) -- MO 2979/1-1.

\section*{Data availability}

The derived data in this article will be shared on reasonable request to the corresponding author.




\bibliographystyle{mnras}
\bibliography{astro} 




\appendix




\bsp	
\label{lastpage}
\end{document}